\documentclass[a4paper,showpacs,twocolumn,floatfix]{revtex4}
\usepackage{graphicx}
\usepackage{amsmath}
\usepackage{amssymb}
\usepackage{amsfonts}
\usepackage{subfigure}

\newcommand{\bra}[1]{\langle{#1}|}
\newcommand{\ket}[1]{|{#1}\rangle}

\begin{document}
\title{Hybrid Luttinger liquid/Fermi liquid behaviour in an atomic
  wire on a surface}

\date{\today} 

\author{L. K. Dash}
\email{louise.dash@ucl.ac.uk}
\author{A. J. Fisher}
\email{andrew.fisher@ucl.ac.uk}
\affiliation{Department of Physics and Astronomy, University College
  London, Gower Street, London WC1E 6BT}

\begin{abstract}
  Recent advances in single atom manipulation have made it possible to
  create ``wires'' in the form of atomic scale linear structures on a
  semiconductor surface.  If such structures are to be used in future
  electronic devices, it will be necessary to know to what extent they
  behave as one-dimensional conductors, in which the presence of
  electron-electron interactions in the wire lead to properties quite
  different from those found in higher-dimensional systems.  In
  particular, it would be useful to know if these structures retain
  any of the Luttinger liquid properties that are predicted for a pure
  one-dimensional metal.  However, experimental studies of such
  structures have so far yielded unclear and contradictory results.
  We investigate this problem theoretically by creating a highly
  simplified model of a wire on a surface---a chain of atoms that
  includes electron-electron interactions to represent the wire,
  coupled to a similar chain that excludes electron-electron
  interactions to represent, albeit crudely, the surface.  We present
  results for the eigenstates and spectral functions for such systems
  that suggest that many Luttinger liquid indicators are present.
  However the system as a whole retains some residual Fermi liquid
  characteristics, and therefore merges properties of both one and
  higher dimensions.
\end{abstract}



\maketitle
\section{Introduction}\label{sec:introduction}

In recent years it has become possible to create atomic scale wires
consisting of a single chain of atoms on a semiconductor surface.
These wires have obvious potential applications in future nanoscale
circuitry, but most of the research effort so far has concentrated on
the electronic structure of these systems.

Moreover, it has been shown that certain of these systems are not
subject to the Peierls transition, even at low temperatures.  Hence
they are expected to behave as one-dimensional metals, with properties
substantially different from the Fermi-liquid properties of
higher-dimensional metals.  Specifically, the discontinuity of the
occupation number at the Fermi energy is lost, spin and charge
excitations become separated and the correlation functions of the
system develop power-law behaviour.  Experimental evidence for such
Luttinger liquid-like behaviour has, however, proved elusive and
ephemeral.  In particular, the case of Au wires on vicinal Si(111) has
attracted much recent discussion
\cite{Segovia:99-I,Altmann:01,Losio:01,Himpsel:01,SanchezPortal:02,Robinson:02},
as have the potential Luttinger liquid properties of
quasi-one-dimensional charge-transfer salts
\cite{Lorenz:02,Claessen:02,Schwartz:98}.

One of the important questions that remains unanswered is to what
extent the coupling of the wire to the surface affects the
one-dimensional properties that would be expected of an isolated wire.
This is a somewhat more complicated problem than that of two coupled
one-dimensional conductors, which has been widely studied
\cite{Capponi:98,Clarke:97,Clarke:97-II,Shannon:97,Capponi:96,Poilblanc:96,Boies:95,Clarke:94,Fabrizio:93,Fabrizio:93-II,Finkelstein:93,Fabrizio:92}

The present work therefore begins to tackle this problem by reducing
it to its most simple form.  The wire is represented as completely as
possible, by a fully interacting one-dimensional conductor based on
the Luttinger model.  This is coupled to a one-dimensional
representation of the surface,  similar to the wire but without any
electron-electron interactions.  In this way we are able to probe the
effects of the coupling of the wire to the surface in as simple a way
as possible, while retaining many of the properties that would be
found in a more realistic system.

In a previous paper \cite{DashFisher01} we described the case where
the one-electron properties of the two chains are the same.  In this
paper we build on these results by treating the more relevant case in
which the properties of the chain differ.  Where relevant, we have
reproduced results from this earlier work.

The remainder of this paper is structured as follows.  We start with a
review of Luttinger liquid theory, followed in section \ref{sec:model}
by a full description of our model and a justification of its choice.
We then present results for the eigenstates of the system, both for
two chains identical with respect to their non-interacting properties,
and for two chains with differing properties.  We then present results
for calculations of the spectral functions for both types of systems,
and calculate the Luttinger parameter $K_\rho$.  The paper concludes
with a discussion of the implications of the results.

\section{Theoretical methods}\label{Theoretical_methods}
\subsection{The Luttinger model}
\label{sec:luttinger-model}

\begin{figure}[htbp]
  \begin{center}
    \includegraphics[height=8cm]{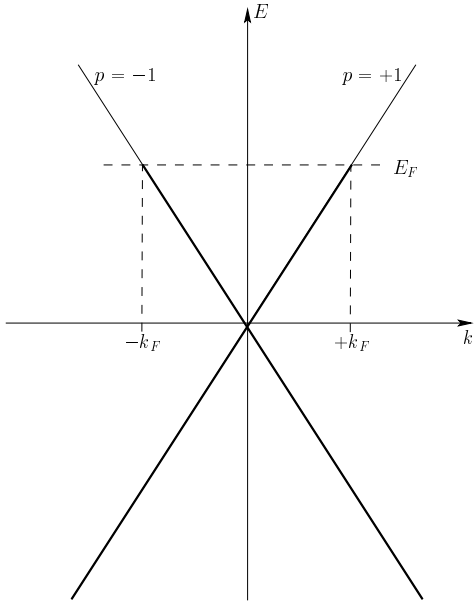}
    \caption{Schematic representation of the Luttinger model.  $p=+1,
      -1$ indicate the populations of left- and right-moving fermions
      respectively. States up to $E_F$ are filled, this includes the
      infinity of negative energy states.}
    \label{fig:Luttinger_model}
  \end{center}
\end{figure}

Our calculations are based on Haldane's solution of the
Tomonaga-Luttinger model \cite{Haldane:81-II}, consisting of a chain
of spinless idealized atoms with periodic boundary conditions.  The
model is characterized by a linearized dispersion relation with slope
$v_{\rm F}$ and a complete separation of the populations of left- and
right-moving particles (as in figure \ref{fig:Luttinger_model}) and
the inclusion of an infinity of negative energy electrons
\cite{Haldane:81-II,Voit:95}. It is this that makes the model exactly
solvable, as the boson commutation relations are then exact rather
than approximate.

The details of the Luttinger model are covered in several review
articles (see, for example, references
\cite{Voit:00,Voit:95,Solyom:79}) and so we shall only review here
those aspects of particular relevance to the current work.  The total
Luttinger model Hamiltonian can be split into three parts
\cite{Voit:95}:
\begin{equation}
  \hat{H} = \hat{H}_0 + \hat{H}_2 + \hat{H}_4. 
\end{equation}$\hat{H}_0$ represents the non-interacting part of the
Hamiltonian:
\begin{equation}
   \hat{H}_0 = \sum_{p,k} v_{\rm F}(pk-k_{\rm F})
   :\hat{c}^\dagger_{pk}\hat{c}_{pk}:,
\end{equation}
where $q$ indicates momentum,$p$ ($=\pm 1$) the branch index and
$:\ldots:$ implies normal ordering.  The interaction parts of the
Hamiltonian, $\hat{H}_2$ and $\hat{H}_4$, are given respectively by
\begin{eqnarray}
  \hat{H}_2 =
   \frac{2}{L}\sum_{q}g_{2}(q)\hat{\rho}_{+}(q)\hat{\rho}_{-}(-q),
   \\
    \hat{H}_4 =
   \frac{1}{L}\sum_{p,q}g_{4}(q)\hat{\rho}_{p}(q)\hat{\rho}_{p}(-q).
\end{eqnarray}
The $\hat{H}_2$ term represents forward scattering between the left-
and right-moving fermion branches, while the $\hat{H}_4$ term
represents forward scattering within a momentum branch.  The density
operators $\hat{\rho}$ are defined in terms of the fermion operators
as
\begin{equation}
  \label{eq:rho}
  \begin{split}
    \hat{\rho}_{p}(q) & = \sum_k
    :\hat{c}^\dagger_{p,k+q}\hat{c}_{p,k}: \\
     & = \sum_{k}\left(\hat{c}^\dagger_{p,k+q}\hat{c}_{p,k} -
      \delta_{q,0}\langle\hat{c}^\dagger_{p,k}\hat{c}_{p,k}\rangle_0
    \right).
\end{split}
\end{equation}

The complete Hamiltonian can be diagonalized via a Bogoliubov transformation
\cite{Haldane:81-II} to yield the bosonized form of the Hamiltonian
\begin{widetext}
\begin{eqnarray}
   \label{eq:bosonized_hamiltonian}
  \tilde{H} = \frac{\pi}{L}\sum_{p,\ q\neq 0} v_0(q)
  :\tilde{\rho}_p(q)\tilde{\rho}_p(-q): +
  \frac{\pi}{2L}\left[v_N\left(N_+ + N_-\right)^2 + v_J\left(N_+ -
      N_-\right)^2\right] \\
   = \frac{1}{2}\sum_q\left(\omega_q -v_{\rm F}|q|\right)+
  \sum_q\omega_q\hat{b}_q ^\dagger\hat{b}_q +
    \frac{\pi}{2L}\left(v_NN^2 + v_JJ^2\right),
\end{eqnarray}
\end{widetext}
where the transformed density operators $\tilde{\rho}$ are related to the originals
by a phase $\phi_q$:
\begin{equation}
  \tilde{\rho}_p(q) = \hat{\rho}_p(q)\cosh \phi_q +
  \hat{\rho}_{-p}(q)\sinh\phi_q.
\end{equation}
There is a characteristic frequency $\omega_q = \sqrt{|(v_{\rm F} +
g_4(q)/2\pi)^2 - (g_2(q)/2\pi)^2|}/|q|$ associated with the
transformed bosons, while the quantum numbers $N \equiv N_+ + N_-$ and
$J \equiv N_+ - N_-$ represent respectively the sum of the number of electrons on
the positive and negative branches, and the difference between them
(analogous to current).

The three velocities in the Hamiltonian are related as follows:
\begin{equation}\label{eq:velocities_1}
  \begin{split}
  v_N v_J &= v_0^2, \\
  v_N &= \frac{v_0}{K_\rho} = v_0  e ^{-2\phi}, \\
  v_J &= v_0 K_\rho = v_0  e ^{+2\phi},
  \end{split}
\end{equation}
and are also related to the non-interacting Fermi velocity by
\begin{equation}\label{eq:velocities_2}
  \begin{split}
    v_N = v_{\rm F} + \frac{g_4 + g_2}{2\pi}, \\
    v_J = v_{\rm F} + \frac{g_4 - g_2}{2\pi}.
    \end{split}
\end{equation}
The requirement that $\tilde{H}$ be a diagonalized version of
$\hat{H}$ is ensured by the relationship between the Bogoliubov
transformation phase $\phi_q$ and the interaction functions $g_i$:
\begin{align}
  \label{eq:K_rho}
   e ^{2\phi_q} &  = \left(\frac{\pi v_{\rm F} + g_4(q) -
  g_2(q)}{\pi v_{\rm F} + g_4(q) + g_2(q)}\right)^{1/2} \\
 & \equiv K_\rho(q).
\end{align}
The spinless Luttinger liquid parameter $K_\rho$ is then obtained by
taking the limit of $K_\rho(q)$ as $q$ tends to zero.  In all of the
above the limit $q\rightarrow 0$ is implied where $q$ is not
explicitly included.

In addition there are two further parameters related to $K_\rho$: for
a spinless Luttinger liquid they are given by
\begin{align}
  \alpha & = \frac{1}{2}\left[K_\rho +
      \frac{1}{K_\rho} -2\right], \label{eq:alpha}\\
    \gamma  & = 2\alpha. \label{eq:gamma}
\end{align}
The Fermi liquid corresponds to $K_\rho = 1$ and $\alpha = 0$ and so
departures from these values can be used to ``measure'' the extent of
non-Fermi liquid behaviour. $\alpha$ is an exponent which governs the
power-law dependence of all single-particle properties (for example
the density of states, which varies as $N(E) \approx
|E-E_{\rm F}|^\alpha$), as well as other properties of the system,
including the d.c. resistivity, which varies as $\rho(T) \approx
T^{1-\gamma}$ \cite{Voit:95,Ogata:92}.

\subsection{The model}\label{sec:model}

We are now able to couple two Luttinger chains (labelled by
superscript $A$  and
$B$) together by allowing hopping between adjacent points on
each chain with matrix element $t_\perp$:
\begin{equation}\label{eq:coupled_hamiltonian}
  \begin{split}
    \hat{H}^{\rm coupled} & = \hat{H}^A + \hat{H}^B \\
    & \qquad +
    t_\perp\sum_{x,p}\left(\hat{\psi}_p^{A\dagger}(x)\hat{\psi}_{p}^{B}(x)
      + \hat{\psi}_{p}^{B\dagger}(x)\hat{\psi}_{p}^{A}(x)\right).
  \end{split}
\end{equation}
Other schemes of inter-chain hopping are of course possible, and we
expect that they will generate similar results.  We neglect any ``drag
effects'' of inter-chain interactions that would be generated by
interaction terms analogous to $\hat{H}_2$ or $\hat{H}_4$ involving
electrons on both chains.  We have the choice of using either a set of
basis states generated by the diagonalized boson operators
$\hat{b}_q^\dagger$ or the non-diagonalized operators
$\hat{a}_q^\dagger$.  We choose the latter, as although this has the
consequence that the ground states for a single interacting chain no
longer consist of the relevant zero-boson basis states, it is
considerably more convenient computationally.  The
$\hat{\psi}^\dagger$ are fermion creation operators given in the
bosonized form by
\begin{equation}
  \label{eq:fermion_boson}
  \hat{\psi}_{p,c}^{\dagger}(x) = \frac{1}{\sqrt{L}}  e ^{ i 
  pk_{\rm F}x} 
  \left[e^{i\hat{\phi}_{p,c}^{\dagger}(x)} \hat{U}_{p,c}
   e^{i\hat{\phi}_{p,c}(x)}\right],
\end{equation}
$\hat{\phi}_{p,c}(x)$ is a boson field operator given by
\begin{equation}
  \label{eq:boson_field_operator}
  \hat{\phi}_{p,c}(x) = \left(\frac{p \pi x}{L}\right)N_{p,c} + 
  i\sum_{q\neq 0}\theta(pq)\left(\frac{2\pi}{L|q|}
  \right)^{1/2} e ^{ i qx}\hat{a}_{q,c},
\end{equation}
where the subscripts $p$ refers to the branch index ($\pm 1$) and $c$
to the chain index.  $\hat{U}_{p,c}$ is a ladder operator whose form
ensures the anticommutation properties of the final fermion operator $
\hat{\psi}_p^{\dagger}(x)$ despite the commutation properties of the
constituent boson operators $\hat{a}_q$.

In order for $\hat{U}_{p,c}$ to produce anticommuting field operators
on different chains, it is necessary to introduce a further phase
factor into its definition, analogous to Haldane's phase factor
$\zeta(p,N_p,N_{-p})$ for ensuring anticommutation between the
branches of a single chain.  The total ladder operator component of
equation (\ref{eq:fermion_boson}) thus takes the form
\begin{equation}\label{eq:ladderoperator}
  \hat{U}_{p,c} =
  \zeta(p,N_p,N_{-p})\zeta^\prime(c,N_c,N_{-c})|N_{p,c}+1, N_{-p,c}\rangle,
\end{equation}
where the subscript $c = \pm 1$ is a chain index. The anticommutation
properties originate in the phase factors $\zeta$, which can be
written as
\begin{equation}\label{phase_factor_zeta}
  \zeta_{i=p,c} = (-1)^{(\frac{1}{2}iN_{-i})}.
\end{equation}

This model provides us with a substrate that is metallic, whereas in
reality we want to calculate for an insulating substrate.  This may be
achieved within the limitations of the Luttinger model in two ways.
The first of these, reported here, is to adjust the Fermi velocity of
the ``substrate'' chain so that the electrons preferentially reside on
the ``wire'' chain.  Our method of achieving this is described in
detail in section \ref{sec:Different chains}. The second approach, to
be reported elsewhere, is to move all the single-electron states
associated with the substrate chain up in energy.  This therefore has
the effect of simulating a gap between the valence and (largely
inaccessible) conduction bands of the substrate chain, while the
states belonging to the wire chain are located within this gap.

\subsection{Computational details}
\label{sec:Computational_details}

The choice of a specific form for the interactions $g_2(q)$ and
$g_4(q)$ is arbitrary, as
the Luttinger model remains completely solvable for any interaction
that fulfils certain conditions \cite{Haldane:81-II}.  We choose a
Gaussian form for the interaction and set $g_2(q) = g_4(q) = 2\pi
V(q)$ with
\begin{equation}
  \label{eq:gaussian_interaction}
  V(q) = I \exp{\left(-2q^2/r\right)}
\end{equation}
as this has the advantage that it maintains the same form in both real
and momentum space. Other forms, such as a screened Coulomb
interaction, would also be possible.  The parameters $I$ and $r$ can
be varied to control the ``strength'' and range of the interaction
respectively.  We complete the basic two chain model by setting the interaction
strength on chain B, representing the substrate, to zero but retaining interactions on chain A.

\begin{figure}
  \begin{center}
    \includegraphics[width=\columnwidth]{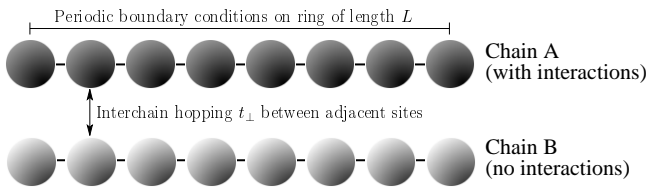}
    \caption{Schematic representation of the two chain model.  Electron-electron
      interactions are included on just one of the chains.}
    \label{fig:chain_hopping}
  \end{center}
\end{figure}

The size of the computational Hilbert space is restricted by allowing
only one boson in each mode.  This does not affect the low energy
properties in which we are interested, as we have checked that only
higher energy excitations involve basis states with more than one
boson in each mode.  We also neglect the presence of the infinity of
negative energy electrons, i.e.
$\hat{c}\ket{N_\textrm{\scriptsize{electrons}}=0} = 0 $.  However, we
are still strongly limited in the size of the system we can handle:
the Hilbert space scales exponentially with both length of chains and
number of electrons.  The largest system for which we have been able
to calculate with arbitrary numbers of electrons is one with length $L
= 6$ (i.e. 12 sites in total).  

\subsection{Spectral functions}
\label{sec:Spectral_functions_theory}
We define the spectral function $\rho(q,\omega)$ as 
\begin{equation}
  \label{eq:spectral_function}
  \rho(q,\omega)_{p,c} = -\frac{1}{\pi} \Im\left[ G^R_{p,c}(k_{\rm
 F} + q, \omega + \mu)\right],
\end{equation}
with $q=k_{\rm F}-k$, and the retarded Green's function $G^R_{p,c}$ defined
as the double Fourier transform of
\begin{equation}
  \label{eq:retarded_greens_function}
   G^R_{p,c}(x,t) = - i \theta(t)\{ \langle \hat{\psi}_{p,c}(x,t)
   \hat{\psi}_{p,c}^\dagger(0,0)\rangle \}
\end{equation}
where the subscripts $p$ and $c$ again indicate branch and chain
indices.  Whereas for a Fermi liquid we would expect $\rho(0,\omega)$
to be a delta function at the Fermi energy, the situation for a
Luttinger liquid is quite different
\cite{Schoenhammer:93,Voit:93,MedenSchonhammer:92}.  Spectral weight
is repelled from the Fermi surface owing to the virtual particle-hole
excitations generated by the inter-branch scattering term $g_2$,
resulting in a broadened peak.  As $q$ is increased the Fermi liquid
spectral function will merely broaden like $q^2$ reflecting the finite
lifetime of electrons away from $E_F$, but for a Luttinger liquid
there is zero spectral weight within a range $\pm v_0 q$ of the Fermi
energy.  In addition, the negative frequency contribution is
suppressed exponentially with $q$, and for a continuum Luttinger
liquid the positive frequency contributions have a power law
dependence.  For $\omega>0$ this is of the form
$\theta(\omega-v_0q)(\omega-v_0q)^{\gamma-1}$ and for $\omega<0$ it is
of the form $\theta(-\omega-v_0q)(-\omega-v_0q)^\gamma$, with $\gamma$
given by equation (\ref{eq:gamma}) \cite{Voit:95}.

However, for our more complex system it is not easy to obtain an
analytic form and so we resort to computational methods.  We calculate
the Green's function
\begin{equation}
  \label{eq:Green's_function_Haydock}
  \begin{split}
    G^R(k,k^\prime,\omega) & = \bra{N}\hat{c}_k \left[\omega - \hat{H} +
      \varepsilon_N \right]^{-1}\hat{c}^\dagger_{k^\prime}\ket{N} \\
    & \qquad +    
    \bra{N} \hat{c}^\dagger_{k^\prime}\ \left[\omega + \hat{H} -
      \varepsilon_N \right]^{-1}\hat{c}_k\ket{N}
\end{split}
\end{equation}
using Haydock's tridiagonal Lanczos-based procedure
\cite{Haydock:80,GolubVanLoan:96}.  In order to ensure convergence, we put $\omega
\rightarrow \omega + i \eta$, where $\eta$ is an imaginary component
of the energy roughly equal to the level spacing of the system
\cite{Haydock:80}. $\ket{N}$ is the $N$-electron ground state, with
energy $\varepsilon_N$.  We also use the Lanczos method to calculate
the eigenstates of the system.

\section{Results}\label{sec:results}

\subsection{Two types of system}
\label{sec:Different chains}

\begin{figure}
  \begin{center}
\subfigure[]{
    \includegraphics[width=6cm]{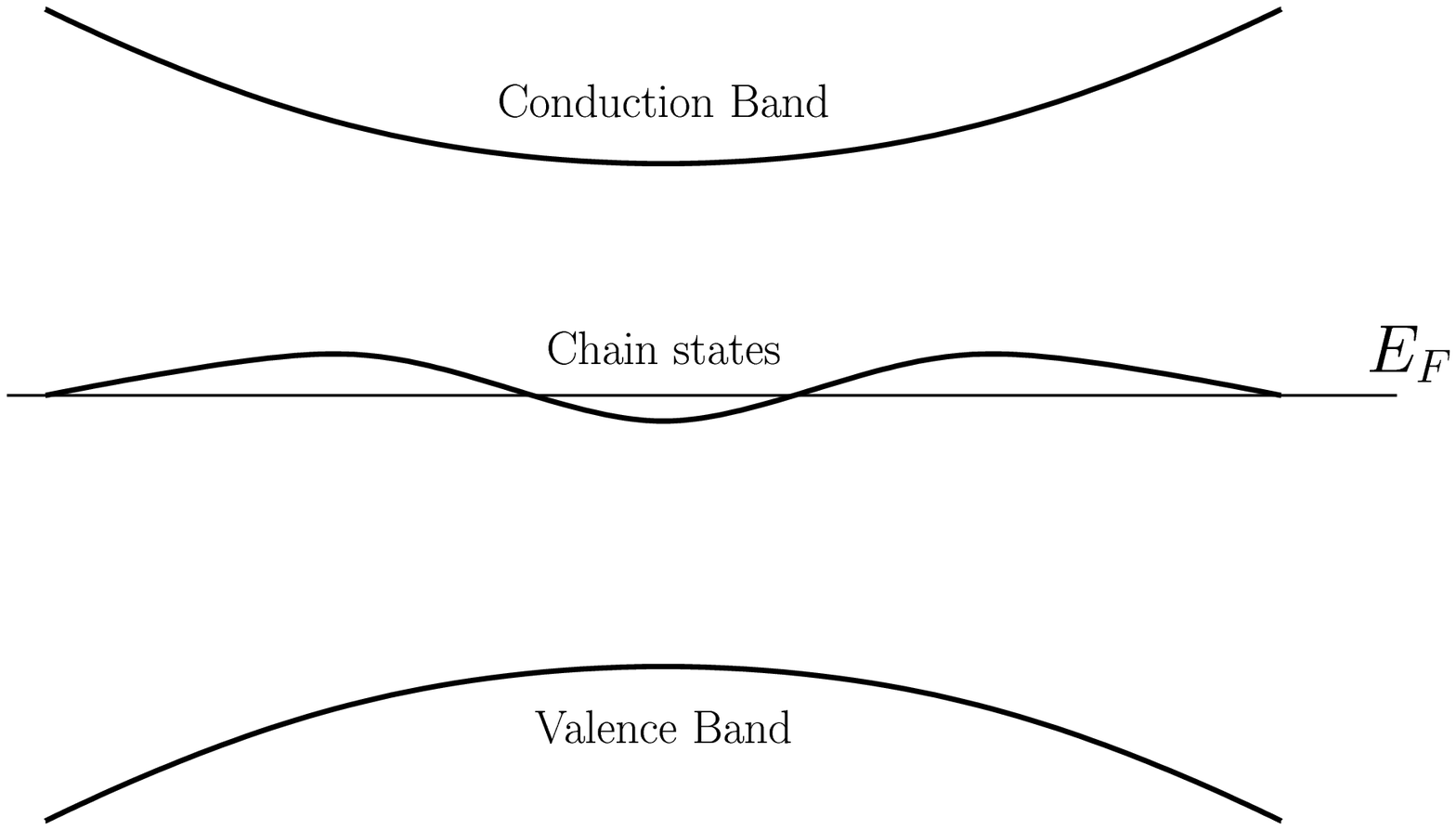}} \\
  \subfigure[]{
    \includegraphics[width=6cm]{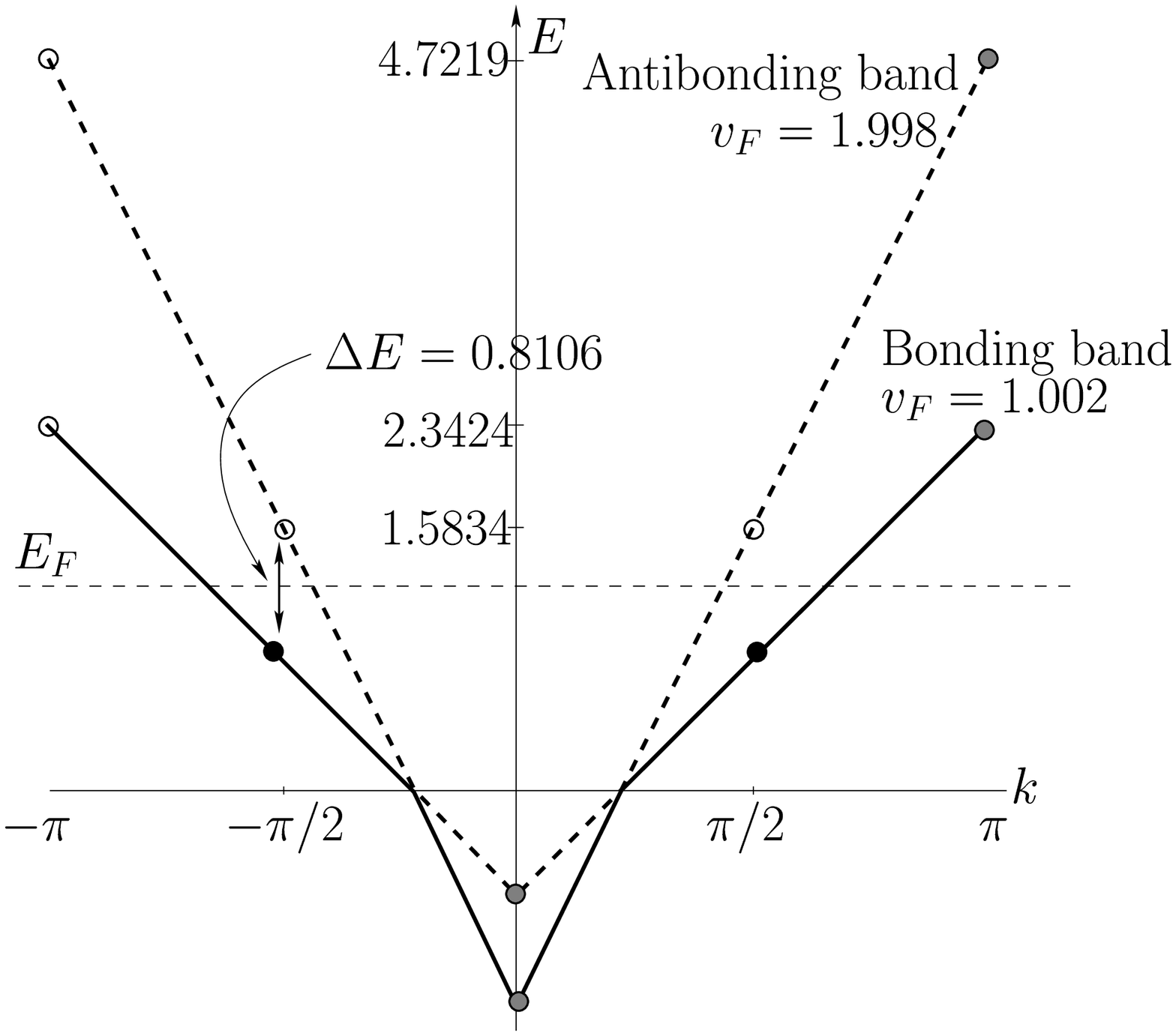}}
  \caption{(a) Schematic band structure of an atomic wire on a
    semiconducting surface, (b) Band structure as realised in our
    calculations for the non-interacting
    system with $L=4$, $N=2$, $t_\perp=0.1$ $v_F^A = 1.0, v_F^B =
    2.0$.  Filled circles represent occupied states, unfilled circles
    empty states.}
  \label{fig:non_ident_bandstructure}
  \end{center}
\end{figure}

Ideally, we would like to be able to calculate for a metallic chain
coupled to an insulating chain representing the substrate.  However,
the metallic nature of our chains is inherent in the Luttinger model
and so this is not possible with the present system.  Nonetheless, we
can reproduce some of the most important characteristics of such a
system within the constraints of the current model.  In particular, we
are interested in systems where the Fermi velocities of the chains
differ, i.e.  the Fermi velocity for the interacting chain is less
than that of the non-interacting chain, $v_F^A < v_F^B$.  The result
of this is that the electrons find it energetically preferable to
reside on the interacting chain.  This is one of the characteristics
we would expect for a metal-semiconductor chain system, where the
electrons in the wire are more mobile than those in the substrate.
This has several consequences for the band structure of the coupled
system, as can be seen more clearly in figure
\ref{fig:non_ident_bandstructure}.  The bands are not parallel and are
not separated by $2t_\perp$, as they are for two chains with equal
$v_F$ The bands anticross at $E=0$, which is $E_F$ for our reference
state of no participating electrons.  Also, for similar chains,
$v_F^A = v_F^B = v_F^{\text{bonding}} = v_F^{\text{antibonding}}$,
whereas in this case we have $v_F^A(=1.0) \approx
v_F^{\text{bonding}}(=1.002)$ and $v_F^B(=2.0) \approx
v_F^{\text{antibonding}}(=1.998)$.

We now present our numerical results for these two systems---the first
with $v_F^A = v_F^B$, the second with $v_F^A < v_F^B$.  The quantities
we calculate are those that may give us information about the
Luttinger liquid properties of the system---specifically, the boson
population of the eigenstates, the excitation energies and
eigenstates, and the spectral function.

\subsection{Boson contributions to eigenstates}
\label{sec:Ground_states}

\begin{figure}[h]
  \begin{center}
    \subfigure[$L=6$, $v_F^A = v_F^B = 1.0$, $t_\perp = 0.1$ and
    $0.5$.  See also reference
    \cite{DashFisher01}.]{\includegraphics[width=\columnwidth]{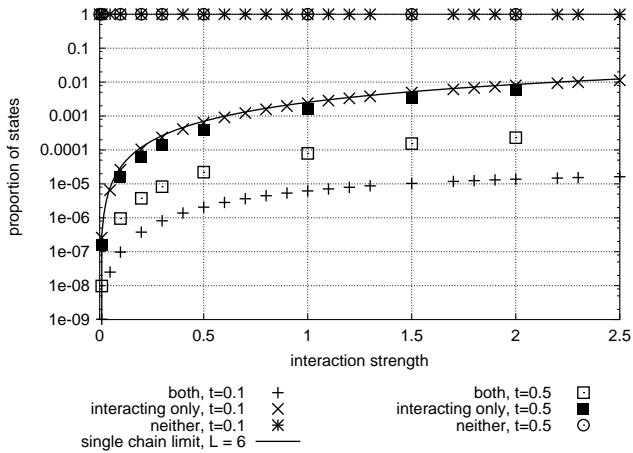}\label{fig:bosoncontributions_a}}
    \\
    \subfigure[$L=4$, $v_F^A = 1.0, v_F^B = 2.0$ and $v_F^A = v_F^B =
    1.0$ ($t_\perp =
    0.1$)]{\includegraphics[width=\columnwidth]{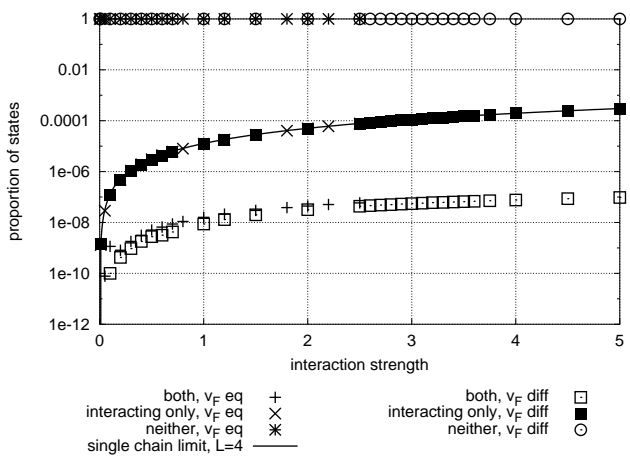}\label{fig:bosoncontributions_b}}
    \caption{Contribution of bosonic basis states to the ground states
      as a function of interaction strength $I$.}
    \label{fig:bosoncontributions}
  \end{center}
\end{figure}

Using the methods detailed above, we calculate the ground and neutral
first excited states for our two systems at various interaction strengths.
We first consider the boson population of these eigenstates.  As we
have chosen to work in the space of un-Bogoliubov transformed bosons,
it is only the non-interacting ($I=0$) systems that have a zero-boson
population in all modes for the ground state.  All the interacting
systems ($I>0$) have contributions from basis states that include
bosons on one or both chains, as these basis states are not
individually eigenstates of the interacting system.  For both our two chain
systems, contributions to the ground state may come from basis states
with no bosons on either chain, bosons on the interacting chain (chain
A) or bosons on both chains.  These contributions are plotted in figure
\ref{fig:bosoncontributions} as a proportion of the total ground
state.  Note that contributions from basis states with bosons on the
non-interacting chain only are negligible and are not included.

For the case where both chains have equal Fermi velocities, as shown
in figure \ref{fig:bosoncontributions_a}, the data for basis states
with bosons only on chain A maps almost exactly onto that for a single
isolated chain.  However, there is also a significant contribution
from states with bosons on both chains which increases with $I$---the
presence of interactions on chain A induces interaction effects on
the otherwise non-interacting chain B.  For stronger interchain
coupling $t_\perp$, we see less of a contribution for both types, but
nonetheless the data for bosonic states on chain A is still comparable
to that of a single chain, indicating that for both $t_\perp = 0.1 $
and $ t_\perp = 0.5$ we are in a weakly coupled regime.

Figure \ref{fig:bosoncontributions_b} compares the results for the
equal $v_F$ system with those from the system with $v_F^A < v_F^B$.
Once more, the data for bosonic basis states only on chain A matches
the single chain data for both of our 2-chain systems.  We again
observe induced chain B boson states at a level similar to that for
the equal $v_F$ system in the $v_F^A < v_F^B$ system.

\subsection{Charge distribution}
\label{sec:Charge_distribution}

\begin{figure}[h]
  \begin{center}
    \subfigure[$L=6$, $v_F^A = v_F^B = 1.0$.  See also reference \cite{DashFisher01}.]{\includegraphics[width=\columnwidth]{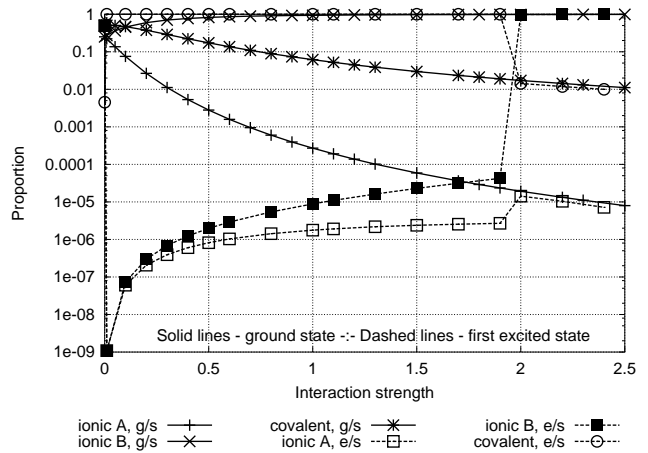}\label{charge_a}}
    \\
    \subfigure[$L=4$, $v_F^A = 1.0, v_F^B = 2.0$]{\includegraphics[width=\columnwidth]{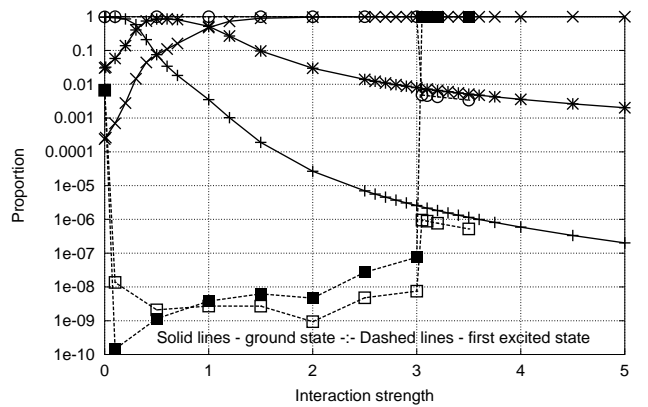}\label{charge_b}}
    \caption{Charge distribution within the chains for the
      ground state and neutral first excited state of the
      two-electron, two-chain model.}
    \label{fig:chargedistribution}
  \end{center}
\end{figure}

We can also examine the distribution of charge between the two chains.
Taking a simple case where there are just two electrons available for
excitation, the possible distributions are both electrons on the
interacting chain, one electron on each chain, or both electrons on
the non-interacting chain.  Those basis states where both electrons
reside on the same chain may be thought of as ``ionic''-like, and
those where the electrons are shared between the chains as
``covalent''-like.  The charge distributions for the ground and first
neutral excited states are shown in figures \ref{charge_a} and
\ref{charge_b}.  We see that for the ground states of the equal $v_F$
system, the eigenstates evolve smoothly from the non-interacting limit
(where the charge distribution is has equal proportions of covalent
and ionic states) to a situation at large values of $I$ where the
ground state is dominated by chain B ionic states.  The presence of
interactions on chain A has thus forced all the available charge onto
the non-interacting chain B.  The case of the excited states is,
however, somewhat different.  The $I=0$ excited state is a fourfold
degenerate combination of ionic and covalent states.  Once even small
chain A interactions are switched on, this degeneracy is lifted and
the dominant basis states are covalent, as indicated by the
discontinuity at $I = 0$ in figure \ref{charge_a}.  As $I$ is
increased, the role of the covalent basis states increases, but there
is a further discontinuity in the charge distribution at $I \approx
2.0$, where the dominance of the covalent states is replaced by that
of chain B ionic states.  This is owing to a change in the nature of
the excitation, from an inter-band excitation with $\Delta k = 0$ to
an intra-band excitation with $\Delta k = 2 \pi v_F / L$.  For $I \geq
2.0$, the charge distribution for the excited state is very similar to
that of the ground state, indicating that no net interchain hopping is
involved in these excitations.

For the $v_F^A < v_F^B$ system, as shown in figure \ref{charge_b}, the
non-interacting ground state is dominated by chain A ionic state, with
lesser contributions from covalent states and a minimal contribution
from chain B ionic states.  As $I$ is increased, these evolve smoothly
through a region ($0.5 \lesssim I \lesssim 1.0$) dominated by covalent
states, to a situation, for $I \gtrsim 1.0$, where the system is
dominated, like the equal $v_F$ case, by chain B ionic states.  The
excited states exhibit a similar pattern of discontinuities to the
equal $v_F$ system, although the transition from inter-band to
intra-band excitations now occurs at a higher interaction strength of
$I = 3.0$.  

\begin{figure}[h]
  \begin{center}
    \subfigure[$L=6$, $v_F^A = v_F^B = 1.0$.  See also reference \cite{DashFisher01}.]{\includegraphics[width=\columnwidth]{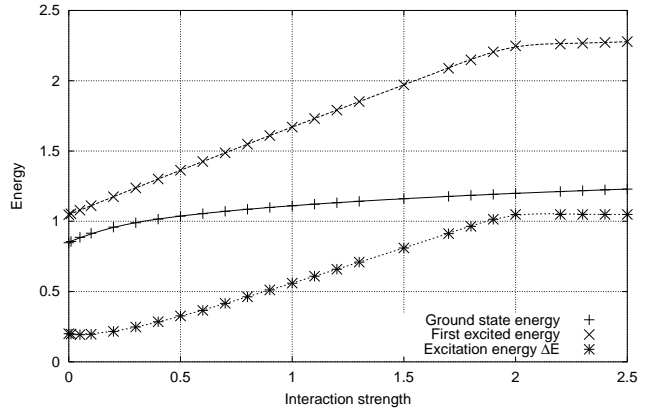}\label{fig:energy_a}}
    \\
    \subfigure[$L=4$, $v_F^A = 1.0, v_F^B = 2.0$]{\includegraphics[width=\columnwidth]{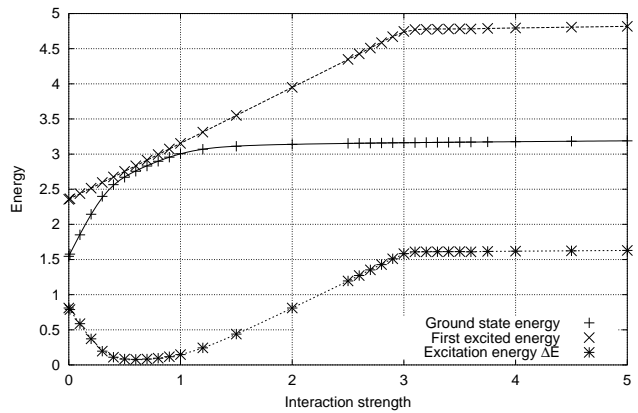}\label{fig:energy_b}}
    \caption{Eigenvalues for the ground state and neutral
      first excited state of the two-electron, two-chain model.}   
    \label{fig:energy_scaling}
  \end{center}
\end{figure}

Having examined the nature of the eigenstates, we now turn our
attention to the eigenvalues, and in particular, to the neutral
excitation energies.  For the equal $v_F$ system, the ground state
energy, first neutral excited state energy and the excitation energy
for various interaction strengths are calculated and shown in figure
\ref{fig:energy_a}.  In the non-interacting limit, the excitation
energy $\Delta E$ is equal to $2t_\perp$, as would be expected for a
transition between bonding and anti-bonding bands.  As $I$ is
increased, $\Delta E$ also increases, until it becomes saturated at $I
\approx 2.0$ at a value of $2 \pi v_F / L$.  This corresponds to the
discontinuity in the charge distribution at $I \approx 2$ (figure
\ref{charge_a}) already associated with the change in nature of the
excitation from inter- to intra-band.

For the system with $v_F^A < v_F^B$, however, the noninteracting
$\Delta E$ is no longer $2t_\perp$ but is approximately $0.81$, as
indicated by the band diagram (figure
\ref{fig:non_ident_bandstructure}).  As $I$ is increased, $\Delta E$
decreases and approaches zero in the range $0.5 \lesssim I \lesssim
1.0$, the same region as is dominated by the covalent states in the
charge distribution.  For $I \gtrsim 1.0$, $\Delta E$ once more increases
until it too becomes saturated at $I \approx 3.0$, at a value of
around 1.6.  This is not, as might be suggested by the fact that all
the charge is on the non-interacting chain, $2 \pi v_F^B / L$, but is
in fact slightly more than the value of $2 \pi v_F^{\text{bonding}} /
L$.  The excitation energy also continues growing for $I > 3.0$,
albeit very slowly: this, and the fact that $\Delta E$ is not exactly
$2 \pi v_F^{\text{bonding}} / L$ at saturation, is because of  the
combined effects
of increase in energy due to induced boson states on chain B and
increase in energy due to chain A interactions overtaking the effect
of charge removal from chain A to chain B.

\subsection{Spectral functions}
\label{sec:Spectral_functions_}

\begin{figure}[h]
  \begin{center}
     \subfigure[$L=6$, $v_F^A = v_F^B = 1.0$]{\includegraphics[width=\columnwidth]{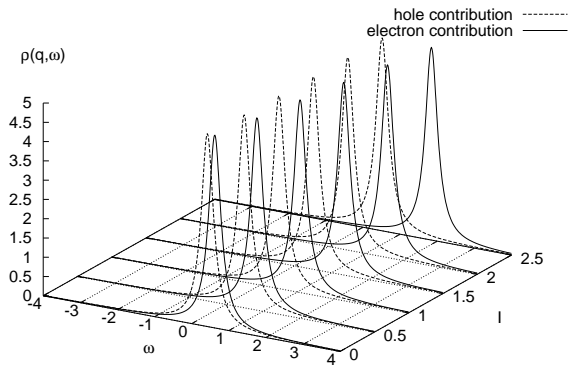}\label{fig:spectral_a}}
    \\
    \subfigure[$L=4$, $v_F^A = 1.0, v_F^B = 2.0$]{\includegraphics[width=\columnwidth]{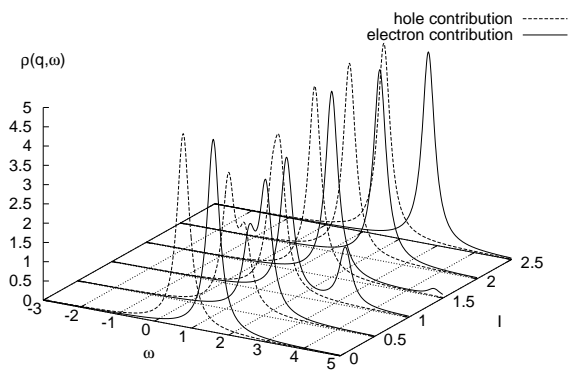}\label{fig:spectral_b}}
    \caption{Spectral functions $\rho(q=0,\omega)$ for both types of
    two chain system. The hole and electron contributions have been
    renormalized to the same height in order to show the positions of
    the peaks accurately.  Creation and annihilation both took place
    on chain A.}
    \label{fig:spectral_functions}
  \end{center}
\end{figure}

The spectral functions for the model are now considered.  Results for
the calculation of the spectral function for both the equal $v_F$ and
the $v_F^A < v_F^B$ systems are presented in figure
\ref{fig:spectral_functions}.  For the equal $v_F$ system, figure
\ref{fig:spectral_a}, there are three sources of broadening of the
peaks away from the delta function we would expect for a Fermi liquid.
The first of these is owing to the finite value of the imaginary
energy $\eta$ used in the numerical calculations, and leads to
broadening of the peaks into Lorentzians with width $\sim \eta$.  The
second results from the splitting into bonding and antibonding bands
due to the interchain coupling and is equal to $2t_\perp$.  However, the
remaining broadening, which is interaction-dependent, is due to the removal of spectral
weight from the Fermi energy at $\omega = 0$, and is indicative of
residual Luttinger liquid behaviour in the coupled system.

For the $v_F^A < v_F^B$ system, figure \ref{fig:spectral_b}, we again see the
broadening of individual peaks into Lorentzians.  However, the peaks
initially converge as $I$ increases, analogous to the similar decrease
in neutral excitation energy observed in the same region of $I
\lesssim 0.5$.  In this region around $I \sim 0.5$ however, two
``branches'' of the spectral function emerge, corresponding to
separate peaks arising from covalent basis states and ionic chain B
basis states---at lower values of $I$ the covalent peak dominates,
whereas for higher values of $I$ the ionic B states become dominant.
For the equal $v_F$ spectral function, a similar analysis reveals that
the single peak is due to ionic B states and there is no corresponding
second peak arising from covalent states.  The peak separation at $I =
0$ is approximately $0.81$, consistent with that indicated by the band
structure, and following the emergence of the two branches noted
above, the peaks separate in the $I \gtrsim 1.0$ region, again
indicating residual Luttinger liquid behaviour.

\subsection{Calculation of the Luttinger parameters}
\label{sec:Luttinger_parameters}

\begin{figure}
  \begin{center}
    \includegraphics[width=\columnwidth]{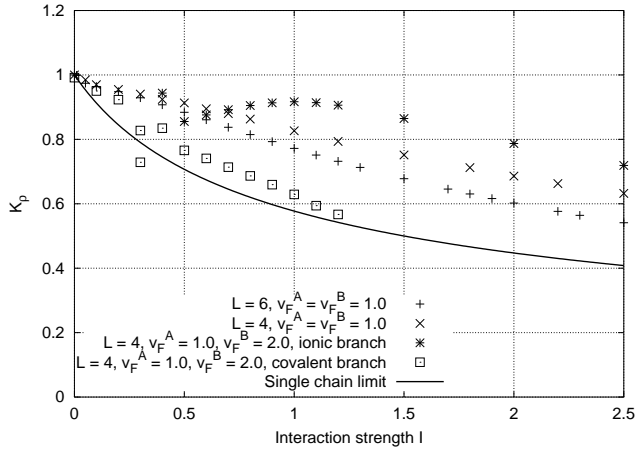}
    \caption{Calculated values of the Luttinger parameter $K_\rho$ for
    several systems.  See also reference \cite{DashFisher01}.}
    \label{fig:luttinger_parameters}
  \end{center}
\end{figure}

We can use the spectral function results to calculate the Luttinger
parameters of the system.  For a continuum system, this could be
achieved by direct numerical measurement of the exponent $\alpha$
\cite{Schoenhammer:93,MedenSchonhammer:92,Voit:93}, but this is not
possible in small discrete systems such as those under present
consideration.  However, we can define an effective Luttinger velocity
$v_0^{\text{eff}}$ for the system as follows.  For a single Luttinger
liquid chain, the separation $\Delta\omega$ of the electron and hole
contributions to the spectral function is equal to $2 v_0 q$, whereas
for a non-interacting double chain system $\Delta\omega$ is equal to
$2 t_\perp$.  We see from figure \ref{fig:spectral_a} that for the
equal $v_F$ system, we have a combination of both these effects, and
hence define our effective $v_0$ in terms of the electron-hole peak
separation
\begin{equation}
  \label{eq:v_0^eff}
  \Delta\omega = 2(v_0^{\text{eff}} + t_\perp).
\end{equation}
As we have chosen $g_2(q)$ to be equal to $g_4(q)$, we can equate
equations \eqref{eq:velocities_1} and \eqref{eq:velocities_2} to give
our effective Luttinger parameter
\begin{equation}
  \label{eq:K_rho^eff}
  K_\rho^{\text{eff}} =  e ^{2\phi} = \frac{v_F}{v_0^{\text{eff}}}.
\end{equation}
The results of this are shown in figure \ref{fig:luttinger_parameters}
and compared to the single-chain continuum limit.

However, for the $v_F^A \neq v_F^B$ system the situation is slightly
different.  In the non-interacting limit the bonding-antibonding
splitting is no longer $2t_\perp$, but instead is of the order of
$q|v_F^A - v_F^B|$.  Since this is, in this case, much larger than
$2t_\perp$, the bonding/antibonding splitting may be ignored,
resulting in the following expression for the effective Luttinger
velocity
\begin{equation}
  v_0^\text{eff} = 
  \begin{cases}
    \frac{\Delta\omega -2t_\perp}{2q} & (v_F^A = v_F^B), \\
    \frac{\Delta\omega}{2q} & |v_F^A - v_F^B|q \gg 2t_\perp.
  \end{cases}
\end{equation}

We may then proceed as before to calculate $v_0^\text{eff}$ and hence
$K_\rho = v_F^\text{\sc system}/v_0^\text{eff}$ for this system.

\begin{figure}[h]
  \begin{center}
    \includegraphics[width=\columnwidth]{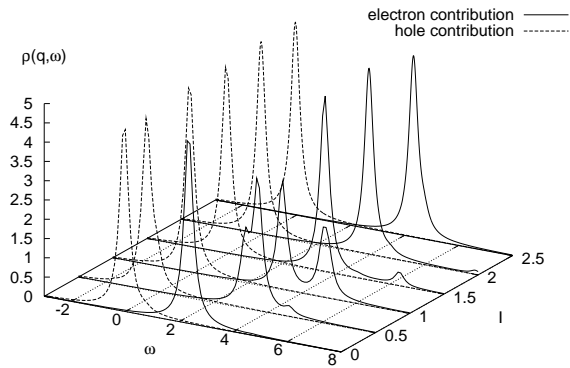}
    \caption{Spectral function $\rho(q=\pi/L,\omega)$ for the $v_F^A =
      1.0$, $v_F^B = 2.0$ two chain system. The hole and electron
      contributions have been renormalized to the same height in order
      to show the positions of the peaks accurately. }
    \label{fig:spectral_fn_qpi}
  \end{center}
\end{figure}

\begin{figure}
  \begin{center}
    \includegraphics[width=\columnwidth]{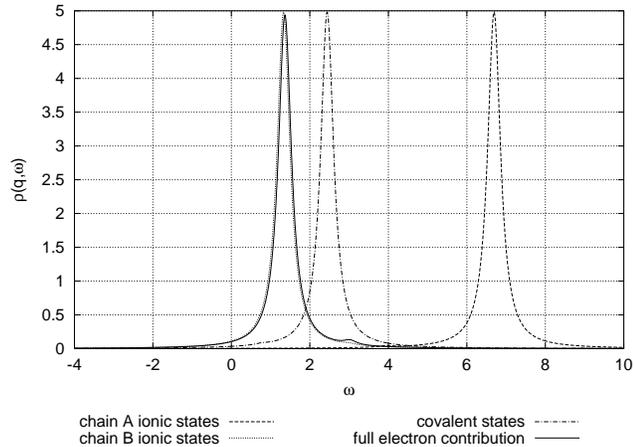}
    \caption{Decomposition of the electron contribution to the
      spectral function into parts due to ionic chain A states, ionic
      chain B states, and covalent states. For the $L=4$, $N=2$,
    $t_\perp=0.1$, $v_F^A = 1.0, v_F^B = 2.0$ system at $I = 0.7$. }
    \label{fig:spectralfn_decomposition}
  \end{center}
\end{figure}

However, we note from figure \ref{fig:spectral_fn_qpi} that there are
two separate peaks in the electron part of $\rho(q,\omega)$.  A
decomposition of the basis states contributing to these states reveals
that the lower energy peak is due to chain B-type ionic states, while
the peak at higher $\omega$ is due to the covalent states, with
virtually no contribution at all from the chain A ionic states. This
is shown in figure \ref{fig:spectralfn_decomposition}. At lower
interaction strengths, the covalent states dominate, but as the
interaction strength increases, the chain B ionic state grow in
importance and the ionic states decrease, fading completely by $I
\approx 2.0$.  The same phenomenon is responsible for the double peak
structure in the hole contribution to $\rho(q=0,\omega)$, where the
effect is somewhat less pronounced.

Either of these peaks can be used to calculate $K_\rho$.  Figure
\ref{fig:luttinger_parameters} shows the results for both the covalent
and ionic peaks, together with the calculated Luttinger parameters for
the $L = 4$ and $L = 6$ equal chain systems.  The values for the
continuum limit of a single chain are also shown for comparison.  We
can see that for $I \lesssim 0.5$, this system is more Luttinger
liquid-like than the equal $v_F$ system, but for interaction strengths
greater than $I \gtrsim 1.0$, the covalent branch peters out and the
system becomes less Luttinger liquid-like than the corresponding equal
chain system.

\section{Conclusions}\label{sec:conclusions}

The basic question we have asked in the current work is
whether the coupling of an Luttinger liquid wire to a surface destroys
any or all of its one-dimensional properties.  We have attempted to
answer this question purely by examination of the eigenstates and
correlation functions of our, albeit simplified, model system.  A more
direct way of answering this question would be to measure the
transport properties directly.  However this would require substantial
changes to the model which we believe would run the risk of obscuring
the basic question: whether one-dimensional metallic systems coupled
to surfaces \emph{in general} retain their one-dimensional properties,
regardless of the influence of such external sources as leads.

\begin{figure}
  \begin{center}
    \includegraphics[width=\columnwidth]{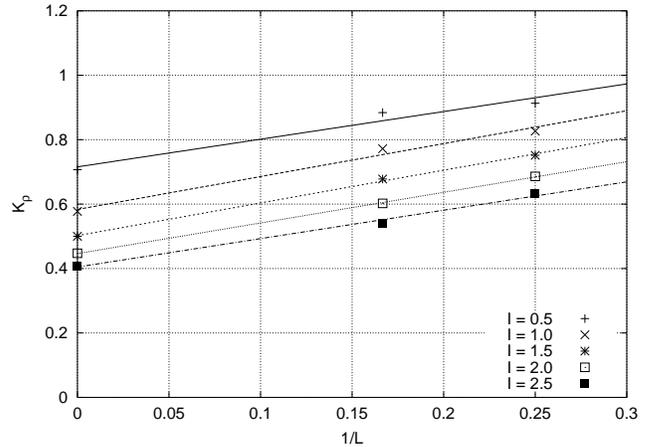}
    \caption{Scaling with respect to inverse system length for the
      calculated values of $K_\rho$ for the $v_F^A = v_F^B = 1.0$
      system at various interaction strengths.  The $1/L = 0$ data are
      for a single-chain continuum system.}
    \label{fig:luttinger_scaling}
  \end{center}
\end{figure}

We have presented results for a system of two chains, identical except
for the presence of electron-electron interactions on one chain,
representing the wire.  We have shown that the coupling to the surface
is weak in nature even for coupling parameters as high as $t_\perp =
0.5$, as the contribution to the ground state from basis states
involving bosons remains comparable to that for an isolated
Luttinger-liquid chain.  We have seen that there is a fundamental
change in the nature of the low-lying neutral excitations at a certain
value of the interaction strength, owing mainly to the finite length
of the chains.  However, this shift, from inter-band excitations with
$\Delta k=0$ to intra-band excitations with $\Delta k= 2\pi v_F/L$ also
leads to a transfer of charge away from the interacting ``wire'' chain
to the non-interacting ``substrate'' chain, resulting in the
restoration of some Fermi liquid properties.  In particular the
excitation energy in this region depends \emph{not} on the Luttinger
velocity $v_0$ but on the Fermi velocity $v_F$.  However, the spectral
properties, which depend on the nature of the charged rather than the
neutral excitations, indicate that Luttinger liquid behaviour survives
at values of the interaction strength not only below this threshold,
but also above it.  This behaviour manifests itself in the removal of
spectral weight from the Fermi surface, and we have used this feature
to calculate effective values of the Luttinger parameter $K_\rho$ for
our system.

Our calculated values of $K_\rho$ indicate definitive Luttinger liquid
behaviour for the system as a whole, and moreover, tend toward the
values for an isolated continuum chain as the system length $L$
increases.  Figure \ref{fig:luttinger_scaling} shows the scaling of
$K_\rho$ as a function of inverse system length for the equal-$v_F$
system at various values of the interaction strength $I$.  The values
for $K_\rho$ for an isolated continuum Luttinger liquid ($L = 0$) are
included and a straight line fit is shown through all three data
points.  One would not expect a perfect straight line fit from these
data---the continuum single chain system remains, after all,
substantially different in nature from a continuum two-chain system.
However, the fact that the data are so linear with respect to inverse
system length further reinforces our general conclusions.

In summary, therefore, we have investigated the problem of
electron-electron interactions in an atomic scale wire on a surface.
We have presented results that indicate that the resulting system does
not behave as a pure Luttinger liquid but nonetheless retains many
characteristics consistent with Luttinger liquid behaviour.  It is
clear from our results, however, that there is an important caveat to
be borne in mind for hybrid systems such as this---that the results
obtained depend very much on the properties of the system one chooses
to measure.

\begin{acknowledgments}
  The authors would like to thank Gillian Gehring for helpful
  discussions and suggestions.  This work was supported financially by
  the UK Engineering and Physical Sciences Research Council and the
  National Physical Laboratory, Teddington, in the form of a CASE
  studentship.  The calculations were performed on the Bentham
  Supercomputer at the HiPerSPACE Computing Centre, UCL, which is
  funded by the UK Engineering and Physical Sciences Research Council.
\end{acknowledgments}


\begin{thebibliography}{32}
\expandafter\ifx\csname natexlab\endcsname\relax\def\natexlab#1{#1}\fi
\expandafter\ifx\csname bibnamefont\endcsname\relax
  \def\bibnamefont#1{#1}\fi
\expandafter\ifx\csname bibfnamefont\endcsname\relax
  \def\bibfnamefont#1{#1}\fi
\expandafter\ifx\csname citenamefont\endcsname\relax
  \def\citenamefont#1{#1}\fi
\expandafter\ifx\csname url\endcsname\relax
  \def\url#1{\texttt{#1}}\fi
\expandafter\ifx\csname urlprefix\endcsname\relax\def\urlprefix{URL }\fi
\providecommand{\bibinfo}[2]{#2}
\providecommand{\eprint}[2][]{\url{#2}}

\bibitem[{\citenamefont{Segovia et~al.}(1999)\citenamefont{Segovia, Purdie,
  Hengsberger, and Baer}}]{Segovia:99-I}
\bibinfo{author}{\bibfnamefont{P.}~\bibnamefont{Segovia}},
  \bibinfo{author}{\bibfnamefont{D.}~\bibnamefont{Purdie}},
  \bibinfo{author}{\bibfnamefont{M.}~\bibnamefont{Hengsberger}},
  \bibnamefont{and} \bibinfo{author}{\bibfnamefont{Y.}~\bibnamefont{Baer}},
  \bibinfo{journal}{Nature} \textbf{\bibinfo{volume}{402}},
  \bibinfo{pages}{504} (\bibinfo{year}{1999}).

\bibitem[{\citenamefont{Altmann et~al.}(2001)\citenamefont{Altmann, Crain,
  Kirakosian, Lin, Petrovykh, Himpsel, and Losio}}]{Altmann:01}
\bibinfo{author}{\bibfnamefont{K.~N.} \bibnamefont{Altmann}},
  \bibinfo{author}{\bibfnamefont{J.~N.} \bibnamefont{Crain}},
  \bibinfo{author}{\bibfnamefont{A.}~\bibnamefont{Kirakosian}},
  \bibinfo{author}{\bibfnamefont{J.-L.} \bibnamefont{Lin}},
  \bibinfo{author}{\bibfnamefont{D.Y.}~\bibnamefont{Petrovykh}},
  \bibinfo{author}{\bibfnamefont{F.J.}~\bibnamefont{Himpsel}}, \bibnamefont{and}
  \bibinfo{author}{\bibfnamefont{R.}~\bibnamefont{Losio}},
  \bibinfo{journal}{Physical Review B} \textbf{\bibinfo{volume}{64}},
  \bibinfo{pages}{035406} (\bibinfo{year}{2001}).

\bibitem[{\citenamefont{Losio et~al.}(2001)\citenamefont{Losio, Altmann,
  Kirakosian, Lin, Petrovykh, and Himpsel}}]{Losio:01}
\bibinfo{author}{\bibfnamefont{R.}~\bibnamefont{Losio}},
  \bibinfo{author}{\bibfnamefont{K.N.}~\bibnamefont{Altmann}},
  \bibinfo{author}{\bibfnamefont{A.}~\bibnamefont{Kirakosian}},
  \bibinfo{author}{\bibfnamefont{J.-L.} \bibnamefont{Lin}},
  \bibinfo{author}{\bibfnamefont{D.Y.}~\bibnamefont{Petrovykh}},
  \bibnamefont{and} \bibinfo{author}{\bibfnamefont{F.J.}~\bibnamefont{Himpsel}},
  \bibinfo{journal}{Physical Review Letters} \textbf{\bibinfo{volume}{86}},
  \bibinfo{pages}{4632} (\bibinfo{year}{2001}).

\bibitem[{\citenamefont{S\'anchez-Portal
  et~al.}(2002)\citenamefont{S\'anchez-Portal, Gale, Garc\'ia, and
  Martin}}]{SanchezPortal:02}
\bibinfo{author}{\bibfnamefont{D.}~\bibnamefont{S\'anchez-Portal}},
  \bibinfo{author}{\bibfnamefont{J.~D.} \bibnamefont{Gale}},
  \bibinfo{author}{\bibfnamefont{A.}~\bibnamefont{Garc\'ia}}, \bibnamefont{and}
  \bibinfo{author}{\bibfnamefont{R.~M.} \bibnamefont{Martin}},
  \bibinfo{journal}{Physical Review B} \textbf{\bibinfo{volume}{65}},
  \bibinfo{pages}{081401} (\bibinfo{year}{2002}).

\bibitem[{\citenamefont{Himpsel et~al.}(2001)\citenamefont{Himpsel, Altmann,
  Bennewitz, Crain, Kirakosian, Lin, and McChesney}}]{Himpsel:01}
\bibinfo{author}{\bibfnamefont{F.}~\bibnamefont{Himpsel}},
  \bibinfo{author}{\bibfnamefont{K.}~\bibnamefont{Altmann}},
  \bibinfo{author}{\bibfnamefont{R.}~\bibnamefont{Bennewitz}},
  \bibinfo{author}{\bibfnamefont{J.}~\bibnamefont{Crain}},
  \bibinfo{author}{\bibfnamefont{A.}~\bibnamefont{Kirakosian}},
  \bibinfo{author}{\bibfnamefont{J.-L.} \bibnamefont{Lin}}, \bibnamefont{and}
  \bibinfo{author}{\bibfnamefont{J.}~\bibnamefont{McChesney}},
  \bibinfo{journal}{Journal of Physics: Condensed Matter}
  \textbf{\bibinfo{volume}{13}}, \bibinfo{pages}{11.097}
  (\bibinfo{year}{2001}).

\bibitem[{\citenamefont{Robinson et~al.}(2002)\citenamefont{Robinson, Bennett,
  and Himpsel}}]{Robinson:02}
\bibinfo{author}{\bibfnamefont{I.K.}~\bibnamefont{Robinson}},
  \bibinfo{author}{\bibfnamefont{P.A.}~\bibnamefont{Bennett}}, \bibnamefont{and}
  \bibinfo{author}{\bibfnamefont{F.J.}~\bibnamefont{Himpsel}},
  \bibinfo{journal}{Physical Review Letters} \textbf{\bibinfo{volume}{88}},
  \bibinfo{pages}{096104} (\bibinfo{year}{2002}).

\bibitem[{\citenamefont{Lorenz et~al.}(2002)\citenamefont{Lorenz, Hoffman,
  Gr\"uniger, Freimuth, Uhrig, Dumm, and Dressel}}]{Lorenz:02}
\bibinfo{author}{\bibfnamefont{T.}~\bibnamefont{Lorenz}},
  \bibinfo{author}{\bibfnamefont{M.}~\bibnamefont{Hoffman}},
  \bibinfo{author}{\bibfnamefont{M.}~\bibnamefont{Gr\"uniger}},
  \bibinfo{author}{\bibfnamefont{A.}~\bibnamefont{Freimuth}},
  \bibinfo{author}{\bibfnamefont{G.}~\bibnamefont{Uhrig}},
  \bibinfo{author}{\bibfnamefont{M.}~\bibnamefont{Dumm}}, \bibnamefont{and}
  \bibinfo{author}{\bibfnamefont{M.}~\bibnamefont{Dressel}},
  \bibinfo{journal}{Nature} \textbf{\bibinfo{volume}{418}},
  \bibinfo{pages}{614} (\bibinfo{year}{2002}).

\bibitem[{\citenamefont{Schwartz et~al.}(1998)\citenamefont{Schwartz, Dressel,
  Gr{\"u}ner, Vescoli, Degiorgi, and Giamarchi}}]{Schwartz:98}
\bibinfo{author}{\bibfnamefont{A.}~\bibnamefont{Schwartz}},
  \bibinfo{author}{\bibfnamefont{M.}~\bibnamefont{Dressel}},
  \bibinfo{author}{\bibfnamefont{G.}~\bibnamefont{Gr{\"u}ner}},
  \bibinfo{author}{\bibfnamefont{V.}~\bibnamefont{Vescoli}},
  \bibinfo{author}{\bibfnamefont{L.}~\bibnamefont{Degiorgi}}, \bibnamefont{and}
  \bibinfo{author}{\bibfnamefont{T.}~\bibnamefont{Giamarchi}},
  \bibinfo{journal}{Physical Review B} \textbf{\bibinfo{volume}{58}},
  \bibinfo{pages}{1261} (\bibinfo{year}{1998}).

\bibitem[{\citenamefont{Claessen et~al.}(2002)\citenamefont{Claessen, Sing,
  Schwingenschl\"ogl, Blaha, Dressel, and Jacobsen}}]{Claessen:02}
\bibinfo{author}{\bibfnamefont{R.}~\bibnamefont{Claessen}},
  \bibinfo{author}{\bibfnamefont{M.}~\bibnamefont{Sing}},
  \bibinfo{author}{\bibfnamefont{U.}~\bibnamefont{Schwingenschl\"ogl}},
  \bibinfo{author}{\bibfnamefont{P.}~\bibnamefont{Blaha}},
  \bibinfo{author}{\bibfnamefont{M.}~\bibnamefont{Dressel}}, \bibnamefont{and}
  \bibinfo{author}{\bibfnamefont{C.S.}~\bibnamefont{Jacobsen}},
  \bibinfo{journal}{Physical Review Letters} \textbf{\bibinfo{volume}{88}},
  \bibinfo{pages}{Art. no 096402} (\bibinfo{year}{2002}).

\bibitem[{\citenamefont{Capponi et~al.}(1998)\citenamefont{Capponi, Poilblanc,
  and Arrigoni}}]{Capponi:98}
\bibinfo{author}{\bibfnamefont{S.}~\bibnamefont{Capponi}},
  \bibinfo{author}{\bibfnamefont{D.}~\bibnamefont{Poilblanc}},
  \bibnamefont{and} \bibinfo{author}{\bibfnamefont{E.}~\bibnamefont{Arrigoni}},
  \bibinfo{journal}{Physical Review B} \textbf{\bibinfo{volume}{57}},
  \bibinfo{pages}{6360} (\bibinfo{year}{1998}).

\bibitem[{\citenamefont{Clarke and Strong}(1997{\natexlab{a}})}]{Clarke:97}
\bibinfo{author}{\bibfnamefont{D.~G.} \bibnamefont{Clarke}} \bibnamefont{and}
  \bibinfo{author}{\bibfnamefont{S.~P.}~\bibnamefont{Strong}},
  \bibinfo{journal}{Journal of Physics: Condensed Matter}
  \textbf{\bibinfo{volume}{9}}, \bibinfo{pages}{3853}
  (\bibinfo{year}{1997}{\natexlab{a}}).

\bibitem[{\citenamefont{Clarke and Strong}(1997{\natexlab{b}})}]{Clarke:97-II}
\bibinfo{author}{\bibfnamefont{D.~G.} \bibnamefont{Clarke}} \bibnamefont{and}
  \bibinfo{author}{\bibfnamefont{S.~P.}~\bibnamefont{Strong}},
  \bibinfo{journal}{Physical Review Letters} \textbf{\bibinfo{volume}{78}},
  \bibinfo{pages}{563} (\bibinfo{year}{1997}{\natexlab{b}}).

\bibitem[{\citenamefont{Shannon et~al.}(1997)\citenamefont{Shannon, Li, and
  d'Abrumenil}}]{Shannon:97}
\bibinfo{author}{\bibfnamefont{N.}~\bibnamefont{Shannon}},
  \bibinfo{author}{\bibfnamefont{Y.}~\bibnamefont{Li}}, \bibnamefont{and}
  \bibinfo{author}{\bibfnamefont{N.}~\bibnamefont{d'Ambrumenil}},
  \bibinfo{journal}{Physical Review B} \textbf{\bibinfo{volume}{55}},
  \bibinfo{pages}{12963} (\bibinfo{year}{1997}).

\bibitem[{\citenamefont{Capponi et~al.}(1996)\citenamefont{Capponi, Poilblanc,
  and Mila}}]{Capponi:96}
\bibinfo{author}{\bibfnamefont{S.}~\bibnamefont{Capponi}},
  \bibinfo{author}{\bibfnamefont{D.}~\bibnamefont{Poilblanc}},
  \bibnamefont{and} \bibinfo{author}{\bibfnamefont{F.}~\bibnamefont{Mila}},
  \bibinfo{journal}{Physical Review B} \textbf{\bibinfo{volume}{54}},
  \bibinfo{pages}{17,547} (\bibinfo{year}{1996}).

\bibitem[{\citenamefont{Poilblanc et~al.}(1996)\citenamefont{Poilblanc, Endres,
  Mila, Zacher, Capponi, and Hanke}}]{Poilblanc:96}
\bibinfo{author}{\bibfnamefont{D.}~\bibnamefont{Poilblanc}},
  \bibinfo{author}{\bibfnamefont{H.}~\bibnamefont{Endres}},
  \bibinfo{author}{\bibfnamefont{F.}~\bibnamefont{Mila}},
  \bibinfo{author}{\bibfnamefont{M.~G.}~\bibnamefont{Zacher}},
  \bibinfo{author}{\bibfnamefont{S.}~\bibnamefont{Capponi}}, \bibnamefont{and}
  \bibinfo{author}{\bibfnamefont{W.}~\bibnamefont{Hanke}},
  \bibinfo{journal}{Physical Review B} \textbf{\bibinfo{volume}{54}},
  \bibinfo{pages}{10,261} (\bibinfo{year}{1996}).

\bibitem[{\citenamefont{Boies et~al.}(1995)\citenamefont{Boies, Bourbonnais,
  and Tremblay}}]{Boies:95}
\bibinfo{author}{\bibfnamefont{D.}~\bibnamefont{Boies}},
  \bibinfo{author}{\bibfnamefont{C.}~\bibnamefont{Bourbonnais}},
  \bibnamefont{and} \bibinfo{author}{\bibfnamefont{A.-M.~S.}
  \bibnamefont{Tremblay}}, \bibinfo{journal}{Physical Review Letters}
  \textbf{\bibinfo{volume}{74}}, \bibinfo{pages}{968} (\bibinfo{year}{1995}).

\bibitem[{\citenamefont{Clarke et~al.}(1994)\citenamefont{Clarke, Strong, and
  Anderson}}]{Clarke:94}
\bibinfo{author}{\bibfnamefont{D.~G.} \bibnamefont{Clarke}},
  \bibinfo{author}{\bibfnamefont{S.~P.}~\bibnamefont{Strong}}, \bibnamefont{and}
  \bibinfo{author}{\bibfnamefont{P.~W.}~\bibnamefont{Anderson}},
  \bibinfo{journal}{Physical Review Letters} \textbf{\bibinfo{volume}{72}},
  \bibinfo{pages}{3218} (\bibinfo{year}{1994}).

\bibitem[{\citenamefont{Fabrizio}(1993)}]{Fabrizio:93}
\bibinfo{author}{\bibfnamefont{M.}~\bibnamefont{Fabrizio}},
  \bibinfo{journal}{Physical Review B} \textbf{\bibinfo{volume}{48}},
  \bibinfo{pages}{15 838} (\bibinfo{year}{1993}).

\bibitem[{\citenamefont{Fabrizio and Parola}(1993)}]{Fabrizio:93-II}
\bibinfo{author}{\bibfnamefont{M.}~\bibnamefont{Fabrizio}} \bibnamefont{and}
  \bibinfo{author}{\bibfnamefont{A.}~\bibnamefont{Parola}},
  \bibinfo{journal}{Physical Review Letters} \textbf{\bibinfo{volume}{70}},
  \bibinfo{pages}{226} (\bibinfo{year}{1993}).

\bibitem[{\citenamefont{Finkelstein and Larkin}(1993)}]{Finkelstein:93}
\bibinfo{author}{\bibfnamefont{A.~M.}~\bibnamefont{Finkelstein}} \bibnamefont{and}
  \bibinfo{author}{\bibfnamefont{A.~I.}~\bibnamefont{Larkin}},
  \bibinfo{journal}{Physical Review B} \textbf{\bibinfo{volume}{47}},
  \bibinfo{pages}{10 461} (\bibinfo{year}{1993}).

\bibitem[{\citenamefont{Fabrizio et~al.}(1992)\citenamefont{Fabrizio, Parola,
  and Tosatti}}]{Fabrizio:92}
\bibinfo{author}{\bibfnamefont{M.}~\bibnamefont{Fabrizio}},
  \bibinfo{author}{\bibfnamefont{A.}~\bibnamefont{Parola}}, \bibnamefont{and}
  \bibinfo{author}{\bibfnamefont{E.}~\bibnamefont{Tosatti}},
  \bibinfo{journal}{Physical Review B} \textbf{\bibinfo{volume}{46}},
  \bibinfo{pages}{3159} (\bibinfo{year}{1992}).

\bibitem[{\citenamefont{Dash and Fisher}(2001)}]{DashFisher01}
\bibinfo{author}{\bibfnamefont{L.K.}~\bibnamefont{Dash}} \bibnamefont{and}
  \bibinfo{author}{\bibfnamefont{A.J.}~\bibnamefont{Fisher}},
  \bibinfo{journal}{Journal of Physics: Condensed Matter}
  \textbf{\bibinfo{volume}{13}}, \bibinfo{pages}{5035} (\bibinfo{year}{2001}).

\bibitem[{\citenamefont{Haldane}(1981)}]{Haldane:81-II}
\bibinfo{author}{\bibfnamefont{F.}~\bibnamefont{Haldane}},
  \bibinfo{journal}{Journal of Physics C} \textbf{\bibinfo{volume}{14}},
  \bibinfo{pages}{2585} (\bibinfo{year}{1981}).

\bibitem[{\citenamefont{Voit}(1995)}]{Voit:95}
\bibinfo{author}{\bibfnamefont{J.}~\bibnamefont{Voit}},
  \bibinfo{journal}{Reports on Progress in Physics}
  \textbf{\bibinfo{volume}{58}}, \bibinfo{pages}{977} (\bibinfo{year}{1995}).

\bibitem[{\citenamefont{Voit}(2000)}]{Voit:00}
\bibinfo{author}{\bibfnamefont{J.}~\bibnamefont{Voit}}, \emph{\bibinfo{title}{A
  brief introduction to {L}uttinger liquids}},
  \bibinfo{howpublished}{cond-mat/0005114} (\bibinfo{year}{2000}).

\bibitem[{\citenamefont{S{\'o}lyom}(1979)}]{Solyom:79}
\bibinfo{author}{\bibfnamefont{J.}~\bibnamefont{S{\'o}lyom}},
  \bibinfo{journal}{Advances in Physics} \textbf{\bibinfo{volume}{28}},
  \bibinfo{pages}{201} (\bibinfo{year}{1979}).

\bibitem[{\citenamefont{Ogata and Anderson}(1993)}]{Ogata:92}
\bibinfo{author}{\bibfnamefont{M.}~\bibnamefont{Ogata}} \bibnamefont{and}
  \bibinfo{author}{\bibfnamefont{P.~W.}~\bibnamefont{Anderson}},
  \bibinfo{journal}{Physical Review Letters} \textbf{\bibinfo{volume}{70}},
  \bibinfo{pages}{3087} (\bibinfo{year}{1993}).

\bibitem[{\citenamefont{Sch{\"o}nhammer and Meden}(1993)}]{Schoenhammer:93}
\bibinfo{author}{\bibfnamefont{K.}~\bibnamefont{Sch{\"o}nhammer}}
  \bibnamefont{and} \bibinfo{author}{\bibfnamefont{V.}~\bibnamefont{Meden}},
  \bibinfo{journal}{Physical Review B} \textbf{\bibinfo{volume}{47}},
  \bibinfo{pages}{16,205} (\bibinfo{year}{1993}).

\bibitem[{\citenamefont{Voit}(1993)}]{Voit:93}
\bibinfo{author}{\bibfnamefont{J.}~\bibnamefont{Voit}},
  \bibinfo{journal}{Journal of Physics C} \textbf{\bibinfo{volume}{5}},
  \bibinfo{pages}{8305} (\bibinfo{year}{1993}).

\bibitem[{\citenamefont{Meden and Sch{\"o}nhammer}(1992)}]{MedenSchonhammer:92}
\bibinfo{author}{\bibfnamefont{V.}~\bibnamefont{Meden}} \bibnamefont{and}
  \bibinfo{author}{\bibfnamefont{K.}~\bibnamefont{Sch{\"o}nhammer}},
  \bibinfo{journal}{Physical Review B} \textbf{\bibinfo{volume}{46}},
  \bibinfo{pages}{15,753} (\bibinfo{year}{1992}).

\bibitem[{\citenamefont{Haydock}(1980)}]{Haydock:80}
\bibinfo{author}{\bibfnamefont{R.}~\bibnamefont{Haydock}},
  \bibinfo{journal}{Solid State Physics} \textbf{\bibinfo{volume}{35}},
  \bibinfo{pages}{215} (\bibinfo{year}{1980}).

\bibitem[{\citenamefont{Golub and van Loan}(1996)}]{GolubVanLoan:96}
\bibinfo{author}{\bibfnamefont{G.~H.} \bibnamefont{Golub}} \bibnamefont{and}
  \bibinfo{author}{\bibfnamefont{C.~F.} \bibnamefont{van Loan}},
  \emph{\bibinfo{title}{Matrix Computations}} (\bibinfo{publisher}{Johns
  Hopkins University Press}, \bibinfo{address}{Baltimore},
  \bibinfo{year}{1996}), \bibinfo{edition}{3rd} ed.

\end{thebibliography}

\end{document}